\documentclass[letterpaper,12pt]{article}
\usepackage{tabularx} 
\usepackage{amsmath,amssymb,amsthm}  
\usepackage{mathrsfs}
\usepackage{enumerate}
\usepackage{float}
\usepackage{indentfirst}
\usepackage{graphicx} 
\usepackage[margin=1in,letterpaper]{geometry} 
\usepackage{caption}
\usepackage{cite} 
\usepackage[colorlinks,linkcolor=magenta,anchorcolor=cyan,citecolor=blue]{hyperref} 
\usepackage{tensor}
\hypersetup{
	colorlinks=true,       
	linkcolor=blue,        
	citecolor=blue,        
	filecolor=magenta,     
	urlcolor=blue
}

\def\be{\begin{equation}}
\def\ee{\end{equation}}
\def\ba{\begin{eqnarray}}
\def\ea{\end{eqnarray}}

 \def\w{\omega}      \def\a {\alpha}    \def\g {\gamma}        \def\b {\beta}      
              \def\.{\cdot}

\begin{document}
\begin{center}

	\vspace{10pt}
	\large{\bf{Spatial and Temporal Chaos of a Bardeen-AdS Black Hole and Effects of Quintessence Dark Energy}}
	
	\vspace{15pt}
	Tian-Zhi Wang , Wen-Biao Liu
	
	\vspace{15pt}
	\small{\it  Department of Physics, Beijing Normal University, Beijing 100875, China}
	
	\vspace{30pt}
\end{center}
\begin{abstract}
   Thermal chaos under spatially/temporally periodic perturbations in the extended phase space of Bardeen-AdS black holes surrounded by quintessence dark energy is investigated.
   The occurring condition of chaos is obtained with the Melnikov integral.
   It is shown that the spatial chaos is always supposed to occur even for a tiny spatially periodic perturbation imposed on the equilibrium configuration.
   However, the temporal chaos will arise in the unstable spinodal region only if the given perturbed amplitude $\g$ is larger than a critical value $\g_c$.
   The value of $\gamma_c$ is determined by the magnetic monopole charge $\beta$, the initial temperature $T_0$, the quintessence state parameter $\omega$, and the quintessence normalization parameter $a$.
   Particularly, combining the effects of $\omega$ and $a$ together, we find that the quintessence appears quite similar to an enhancing/damped mechanism.
   In other words, there exists a critical value $\rho_c$ of the quintessence dark energy density $\rho$.
   In the region of $\rho<\rho_c$, the existence of quintessence leads to a reduction in the viscosity of black holes and thus makes the system more likely to exhibit chaotic behavior.
   Conversely, given the energy density $\rho>\rho_c$, the system acquires higher viscosity so that it is endowed with the ability of enduring a larger thermal fluctuation.
\end{abstract}
	\vfill {\footnotesize ~\\  201921140018@mail.bnu.edu.cn \\ Corresponding author: \\  wbliu@bnu.edu.cn}
\newpage

\section{Introduction}
Chaos, a sort of deterministic but unpredictable phenomenon described by nonlinear or multiple degrees of freedom coupled equations\cite{Hirsch:2013}, is a ubiquitous sight and strange enough to trigger a heated debate both mathematically and physically.
Particularly, as an inherent nonlinear theory, General Relativity(GR) is proved to possess more chaotic properties than the usual dynamical systems.
In the past few decades, various methods have been applied in succession to investigate the chaoticity in the context of geodesic motion of moving objects or some inhomogeneous cosmological models \cite{Bombelli:1991eg,Letelier:1996he,Hanan:2006uf,Witzany:2015yqa,Santoprete:2001wz,Kao:2004qs,Polcar:2019kwu,DeFalco:2021xyo,Letelier:1999xw,Wang:2016wcj,Chen:2016tmr,Liu:2018bmn,Polcar:2019wfi}.
Benefited from these pioneering studies, people do not only acquire more interesting knowledge from the gravitational perspective,
but also strengthen the availability of the Poincar\'{e}-Melnikov approach when they address the dynamical issues of a completely integrable system under some periodic perturbations analytically \cite{Melnikov1963s}.

Not restricted in the framework of nonlinear dynamics, the Melnikov method has also been used to detect the chaotic phenomena arising in a van der Waals fluid system \cite{Polcar:1985}.
It has been shown that under a temporally periodic fluctuation imposed in the unstable spinodal region, the van der Waals system exhibits chaotic behavior only if the given amplitude is larger than a critical value which is relative to a small viscosity.
Besides, the existence of spatial chaos due to a spatially periodic thermal perturbation has also been discovered in the equilibrium configuration even under a tiny perturbation.

Since the prominent contribution of Bekenstein\cite{Bekenstein:1973ur} and Hawking \cite{Hawking:1974sw}, the black hole thermodynamics has been constructed and investigated deeply.
Recently, inspired by the AdS/CFT correspondence \cite{Maldacena:1997re}, people are gradually realizing the significant role of black hole thermodynamics to describe the dual conformal boundary field theory.
Instructively, Refs. \cite{Kastor:2009wy,Kubiznak:2012wp} introduced the concept of black hole extended phase space where one regards the cosmological constant as the thermodynamic pressure $P=-\Lambda/8\pi$ and its conjugate quantity is defined as the thermodynamic volume $V$.
From this point of view, the $P-v$ criticality of Reissner-Nordström-AdS (RN-AdS) black hole extended phase space is, to a large extent, similar to that of van der Waals gas/liquid systems.
Based on this remarkable discovery, multifarious types of the black hole phase transition have been subsequently found to be analogous to the ordinary fluid systems, such as reentrant phase transition (RPT) \cite{Altamirano:2013ane}, triple point \cite{Wei:2021krr} and anomalous van der Waals phase transition \cite{Dehghani:2020kbn}, which forms a new kind of subject named as black hole chemistry.

In view of this great similarity between the extended phase space and the van der Waals system, it is natural to ask if the above-mentioned approach could also be generalized to the black hole system.
In Ref.\cite{Chabab:2018lzf}, the Melnikov method have been successfully applied to test the chaotic behaviors in the RN-AdS black hole extended phase space.
Similarities and differences on thermal chaos and the corresponding critical value are pointed out.
Subsequently, other important generalizations have been carried out in the context of charged Gauss-Bonnet-AdS Black holes\cite{Mahish:2019tgv}, Born-Infeld-AdS black holes\cite{Chen:2019bwt} and charged dilaton-AdS black holes\cite{Dai:2020wny}.
Among them, the dependence of the critical amplitude $\gamma_{c}$ on electric charge $q$, dimension $n$, and other parameters have been investigated.

In this paper, we will generalize these works by calculating the Melnikov functions of the spatially/temporally periodic perturbation in the extended phase space of Bardeen-AdS black holes surrounded by quintessence dark energy.
Not only do we illustrate the dramatic difference between the quintessential Bardeen-AdS case and RN-AdS due to the non-linear electrodynamics, but also we attempt to reveal the effects of quintessence.

On one hand, the regular Bardeen black hole, first proposed in Ref. \cite{AyonBeato:2000zs}, is a spherically symmetric solution resulting from Einstein field equations coupling with a non-linear electromagnetic field.
It satisfies weak energy condition and avoids singularity at the center but preserves an event horizon.
Physical analysis of the matter and thermodynamical variables has shown that the Bardeen model is a fantastic candidate for exploring astrophysical black holes \cite{Bambi:2014nta,Gedela:2021vvu} and a natural particle accelerator \cite{Ghosh:2015pra}.
Therefore, on the basis of the interesting Bardeen model research background, it is significant to detect how the non-linear electrodynamics affects the spacetime, which could provide us with an alternative approach to the combination of GR and quantum theory.

On the other hand, the accelerating expansion of the universe implies a valuable presence of state with the negative pressure.
The origin of the negative pressure could be twofold: cosmological constant and dark energy.
As one of the most suitable candidates for dark energy, the quintessence satisfies the state equation $p_{q}=\omega\rho_{q}$ in the range of $-1<\omega<-1/3$.
Utilizing the phenomenological approach proposed by Kiselev \cite{Kiselev:2002dx}, one can effortlessly and directly be capable of exploring a black hole surrounded by quintessence.
In Refs.\cite{Ghaderi:2017yfr,L.:2018jss,K.:2020rzl}, quintessential Bardeen-AdS black holes have also been investigated successively.
Based on these works, we attempt to test the chaotic behaviors in the extended phase space to find out how the quintessence exactly affects the black holes inside.

The paper is organized as follows.
In Section \ref{sec2}, we give a brief introduction to the thermodynamic setup of the Bardeen-AdS black hole surrounded by quintessence dark energy.
In Section \ref{sec3}, we detect the thermal chaos under a spatially periodic perturbation on the equilibrium configuration with Poincar\'{e}-Melnikov approach.
In Section \ref{sec4}, we switch on the temporally periodic perturbation in the spinodal region and discuss the existing conditions of the thermal chaos.
We draw some conclusions and outlooks of follow-up work in Section \ref{sec5}.
We are here working in Planck units with $\hbar=c=G=k_B=1$.

\section{Thermodynamics of Bardeen-AdS Black Holes Surrounded by Quintessence Dark Energy}\label{sec2}

The Bardeen black hole is a regular spacetime solution of the Einstein's field equation coupled to a non-linear electromagnetic field.
The action in the four-dimensional asymptotically AdS spacetime is given by \cite{AyonBeato:2000zs}
\begin{equation}\label{1}
    \mathcal{S}_{B}=\int {\rm d}^{4}x \sqrt{-g} \left(\frac{R}{16\pi}+\frac{3}{8\pi l^{3}}+\mathcal{L}_Q-\frac{1}{4\pi}\mathcal{L}(\mathcal{F})\right)\,,
\end{equation}
where $l$ is the AdS radius. The first term in parentheses represents the Einstein gravity, the second term denotes the negative cosmological constant $\Lambda=-3l^{-2}$, the third term is contributed by the quintessence dark energy and the forth term is the Lagrangian for a nonlinear electrodynamics source as
\begin{equation}\label{2}
    \mathcal{L}(\mathcal{F})=\frac{3M}{\beta^{3}}\left(\frac{\sqrt{4\beta^{2}\mathcal{F}}}{1+4\beta^{2}\mathcal{F}}\right)^{5/2}\,.
\end{equation}
Here $\mathcal{F}\equiv F^{ab}F_{ab}$, and $\boldsymbol{F}=d \boldsymbol{A}$ is the electromagnetic field strength.
The quantity $\beta$ is the positive magnetic monopole charge.
In the weak field limit $\mathcal{F}\to0$, the nonlinear electrodynamics of the Lagrangian density degenerates to $\mathcal{L}\sim \beta^{-1/2}\mathcal{F}^{5/4}$, which is slightly stronger than Maxwell field \cite{Fan:2016hvf}.

Varying the action (\ref{1}) with respect to the metric tensor $g_{\mu\nu}$ and the electromagnetic field $\mathcal{F}$ respectively, one obtains the equations of motion
\begin{equation}\label{3}
	\begin{aligned}
		G_{\mu\nu}-\frac{3}{l^2}g_{\mu\nu}=&2\left(\frac{\partial \mathcal{L}(\mathcal{F})}{\partial \mathcal{F}}F_{\mu\rho}F_{\nu}^{\,\,\rho}-g_{\mu\nu}\mathcal{L}(\mathcal{F})\right)+T_{\mu\nu}\,,\\
		0=&\nabla_{\mu}\left(\frac{\partial \mathcal{L}(\mathcal{F})}{\partial \mathcal{F}}F^{\mu\nu}\right)\,,
	\end{aligned}
\end{equation}
with the energy momentum tensor $T_{\mu\nu}$ constructed in the Kislev's phenomenological model \cite{Kiselev:2002dx} as 
\begin{equation}
  T_r^r=T_r^r=\rho_{q}\,,\,\,\,\,\,\,\,\,\,\,\,\,\,\,\,\,\,\,\,T_{\theta}^{\theta}=T_{\phi}^{\phi}=-\frac{1}{2}\rho_q\left(3\omega+1\right)\,,
\end{equation}
where $\rho_q$ denotes the energy density of the quintessence $\rho_{q}=-3a\omega/2r^{3\left(\omega+1\right)}$, $\omega$ is the quintessence state parameter within the range $-1<\omega<-1/3$ and $a$ is a positive normalization factor.

Solving the equation of motion Eq.(\ref{3}), a class of static spherically symmetry solutions describing a Bardeen-AdS black hole surrounded by quintessence is \cite{AyonBeato:2000zs}
\begin{equation}
  \begin{aligned}
      ds^2=&-f_{QB}\left(r\right){\rm d}t^2+\frac{1}{f_{QB}\left(r\right)}{\rm d}r^2+r^2{\rm d}\theta^2+r^2\sin^2\theta {\rm d}\phi^2\\
      =&-\left(1-\frac{2Mr^2}{\left(r^2+\beta^2\right)^{3/2}}+\frac{r^2}{l^2}-\frac{a}{r^{3\omega+1}}\right){\rm d}t^2+\left(1-\frac{2Mr^2}{\left(r^2\beta^2\right)^{3/2}}+\frac{r^2}{l^2}-\frac{a}{r^{3\omega+1}}\right)^{-1}{\rm d}r^2\\
      &\,\,+r^2{\rm d}\theta^2+r^2\sin^2\theta {\rm d}\phi^2\,,
  \end{aligned}
\end{equation}
where $M$ is just the Kormar mass originated from the self-gravitating of nonlinear magnetic monopole.

Considering the thermodynamics of the Bardeen-AdS black hole in the extended phase space, the role of the cosmological constant is regarded as a thermodynamic pressure via
\begin{equation}
    P=-\frac{\Lambda}{8\pi}=\frac{3}{8\pi l^{2}}\,.
\end{equation}
One can also calculate Hawking temperature of the Bardeen-AdS black hole with
\begin{equation}\label{7}
    T=\frac{1}{4\pi} \frac{{\rm d}f_{BQ}\left(r\right)}{{\rm d}r}\bigg|_{r=r_h}=\frac{8\pi P r_h^3-2\beta^2 r_h^{-1}+r_h+3a\omega r^{-3\omega}_{h}+3a\beta^2\left(1+\omega\right)r^{-3\omega-2}_{h}}{4\pi\left(r_h^2+\beta^2\right)}\,.
\end{equation}

A series of $P-v$ criticality studies of black holes chemistry have pointed out that it is the horizon radius $r_h$ rather than the thermodynamic volume $V$ that should be associated with the “fluid” volume \cite{Altamirano:2013ane,Wei:2021krr,Dehghani:2020kbn,Ghaderi:2017yfr,L.:2018jss,K.:2020rzl,Kubiznak:2012wp}.
So we abide by the same convention to define the specific volume $v=2r_{h}$.
Then rearranging Eq.(\ref{7}), one can express the equation of state as
\begin{equation}
  P(v,T)=\frac{-v^{3\omega+3}+2\pi T v^{3\omega+4}+8\beta^2v^{3\omega+1}+8\pi\beta^2 T v^{3\w+2}-3\times2^{3\omega+1}a\omega v^2-3\times8^{\omega+1}\beta^2a(1+\omega)}{2\pi v^{-3\omega-5}}\,.
\end{equation}

\begin{figure*}
    \begin{center}
    \includegraphics[width=0.42\textwidth,height=0.21\textheight]{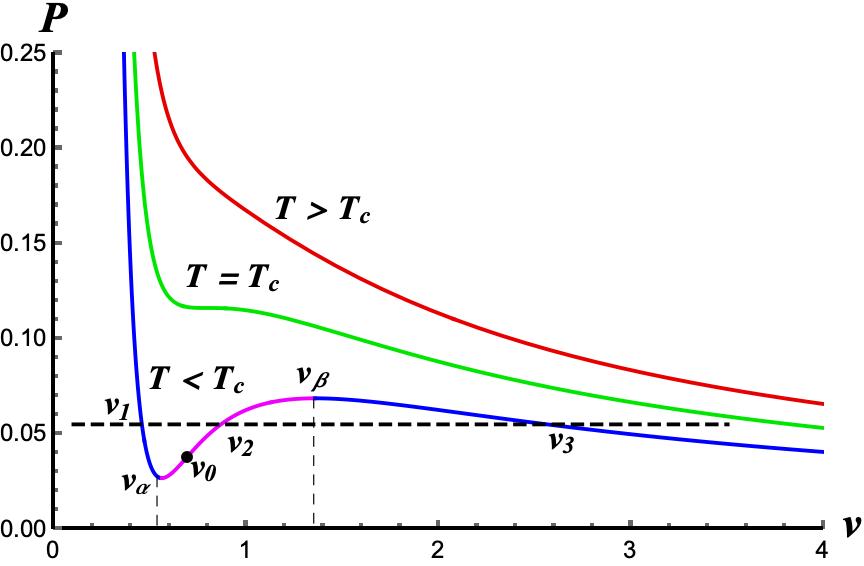}
    \caption{$P-v$ criticality of the Bardeen-AdS black hole surrounded by quintessence at $T>T_c$, $T=T_c$ and $T<T_c$. The $T<T_{c}$ isotherm (the blue curve) is divided into three parts. The middle part (the pink line) of the $T<T_{c}$ isotherm is the spinodal region which contains a saddle point $v_{0}$. The dark dashed line given by the Maxwell's equal area law is the coexisting line of the large and the small black hole phases.\label{fig2.1}}
    \end{center}
\end{figure*}

The $P-v$ isotherm is plotted in Fig.\ref{fig2.1}.
It is shown that there exists a first order phase transition between large black holes and small black holes only if the Hawking temperature $T$ is smaller than a critical value $T_c$ that can be solved by the equations ${\rm d}P (v,T)/ {\rm d} v=0$ and ${\rm d}^2 P(v,T)/{\rm d}v^2=0$.

\section{Spatial Chaos on the Equilibrium Configuration}\label{sec3}
According to the Relativistic Heavy Ion Collider (RHIC) announcement \cite{RHIC}, the quark-gluon-plasma (QGP) does not look like a gas in the free limit but behaves more like a perfect fluid, which means the QGP has a small shear viscosity and thus is strongly-coupled.
Excitingly, this viscosity can be 
successfully predicted by asymptotically AdS black holes in the framework of AdS/CFT correspondence \cite{Maldacena:1997re}.
Moreover, a series of researches on black holes chemistry also imply that black hole system might possess many similar properties as the fluid system both thermodynamically and hydrodynamically.
Therefore, it is natural for us to assume that the quintessential Bardeen-AdS black hole in the extended phase space should act as a thermally compressible and isotropic fluid with a small viscosity.
(Hereafter, this imaginary fluid is called as QB-fluid.)
We limit the QB-fluid into a tube with unit cross section for simplicity.
So the system depends only on the time coordinate $t$ and a Cartesian coordinate $x$ parallel to $r$-direction.
Setting $x_0$ as a reference point, we can write down the relation between the mass and the line density $\rho(x,t)$ of the QB-fluid as
\begin{equation}\label{9}
    M=\int_{x_0}^{x}\rho\left(x',t\right){\rm d}x'\,.
\end{equation}
Considering the QB-fluid with a total mass $2\pi/s$ in the finite tube $2\pi v_0/s$ ($s$ is a positive parameter), we exert a small spatially periodic perturbation of the form
\begin{equation}\label{10}
  T=T_0+\epsilon\cos\left(s x\right)\,,
\end{equation}
where $0<\epsilon\ll 1$.
The initial temperature $T_0$ should be lower than the critical temperature $T_c$ and not too close to $T_c$ to avoid that the phase transition might disappear during the perturbation.
Due to the given perturbation, the previous equilibrium is disturbed and the density of the QB-fluid varies.
Hence, we need a non-linear smooth constitutive equation of stress tensor to respond to this density gradient.
Fortunately, the van der Waals-Korteweg theory provides us with a powerful method to describe this phenomenon of capillarity, to wit the Cauchy tensor $\mathbf{T}$ defined in Ref.\cite{Widom:1977}.
This stress tensor introduces a correction to classical hydrostatics for equilibrium of the compressible QB-fluid and thus avoid the discontinuous jump condition assumed in the classical theory before.
Precisely we write
\begin{equation}
  \mathbf{T}=-P\left(v,T\right)-A\frac{{\rm d}^2v}{{\rm d}x^2}\,,\label{28}
\end{equation}
where $A$ is a positive constant.
There are no body forces in the equilibrium configuration.
So one can obtain $\frac{{\rm d}\mathbf{T}}{{\rm d}x}=0$ which yields the relation $\mathbf{T}=-B$. 
Here, the quantity $B$ can be interpreted as the ambient pressure at both ends of the tube.

\begin{figure*}[h]
  \begin{center}
\includegraphics[width=0.32\textwidth,height=0.18\textheight]{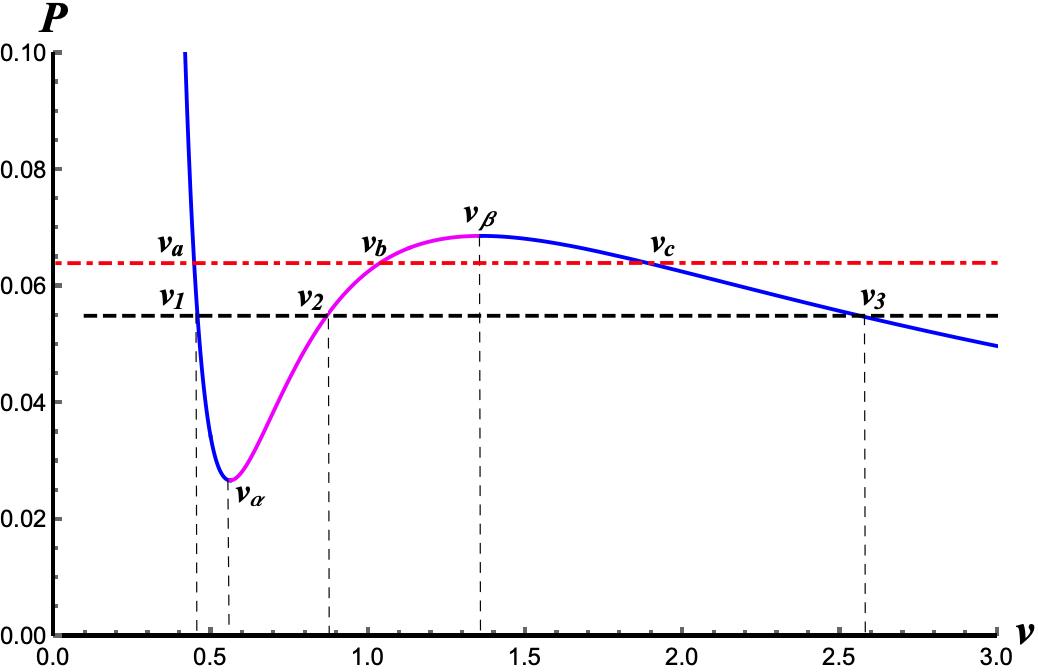}
\includegraphics[width=0.32\textwidth,height=0.18\textheight]{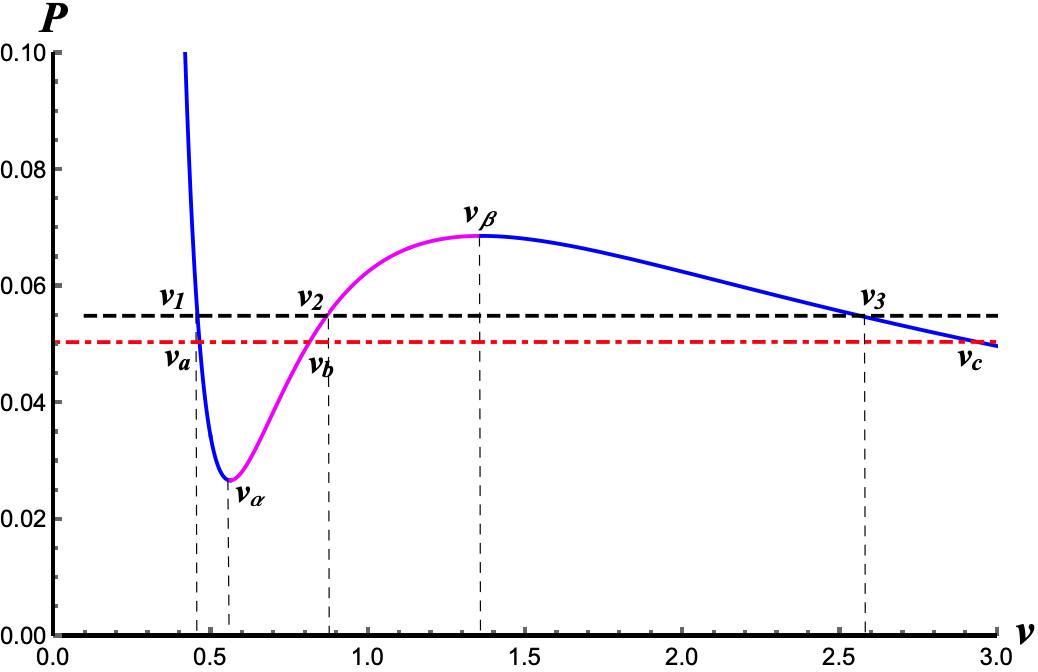}
\includegraphics[width=0.32\textwidth,height=0.18\textheight]{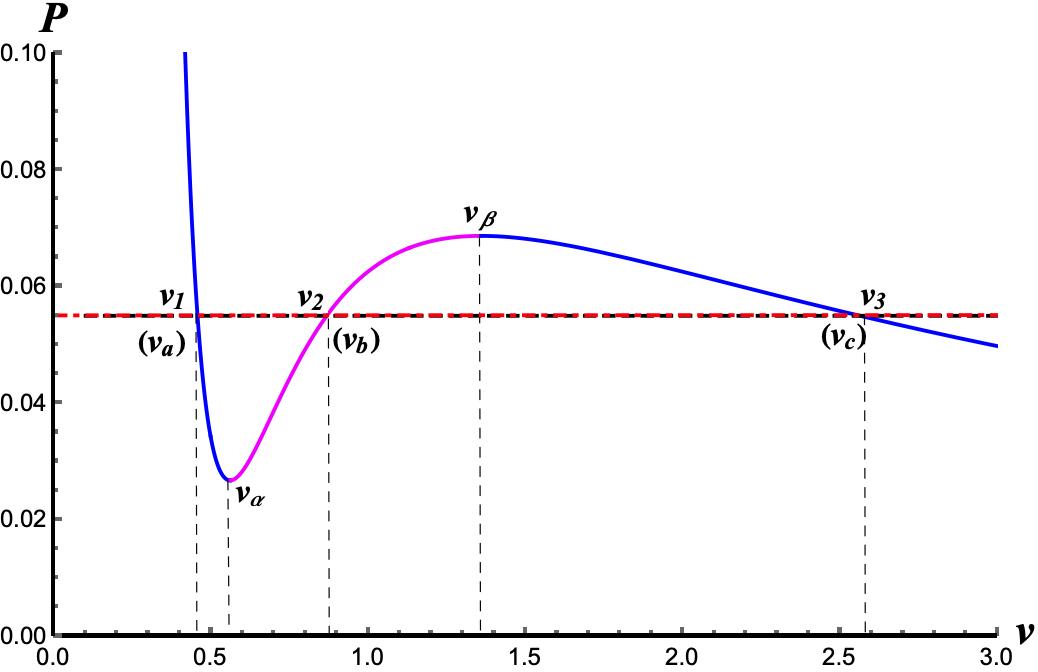}
\includegraphics[width=0.32\textwidth,height=0.18\textheight]{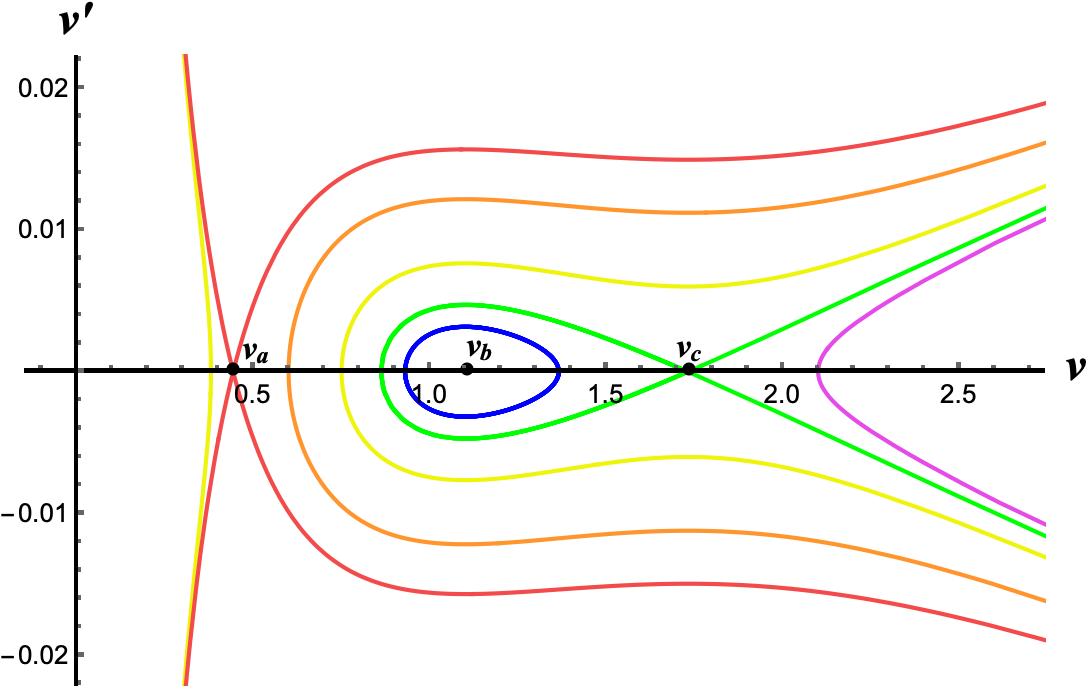}
\includegraphics[width=0.32\textwidth,height=0.18\textheight]{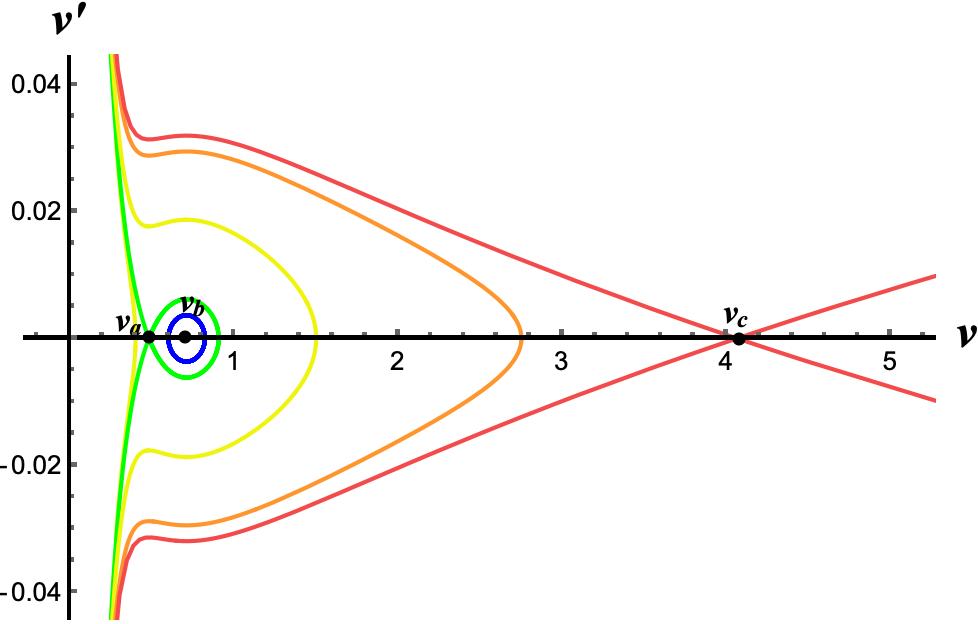}
\includegraphics[width=0.32\textwidth,height=0.18\textheight]{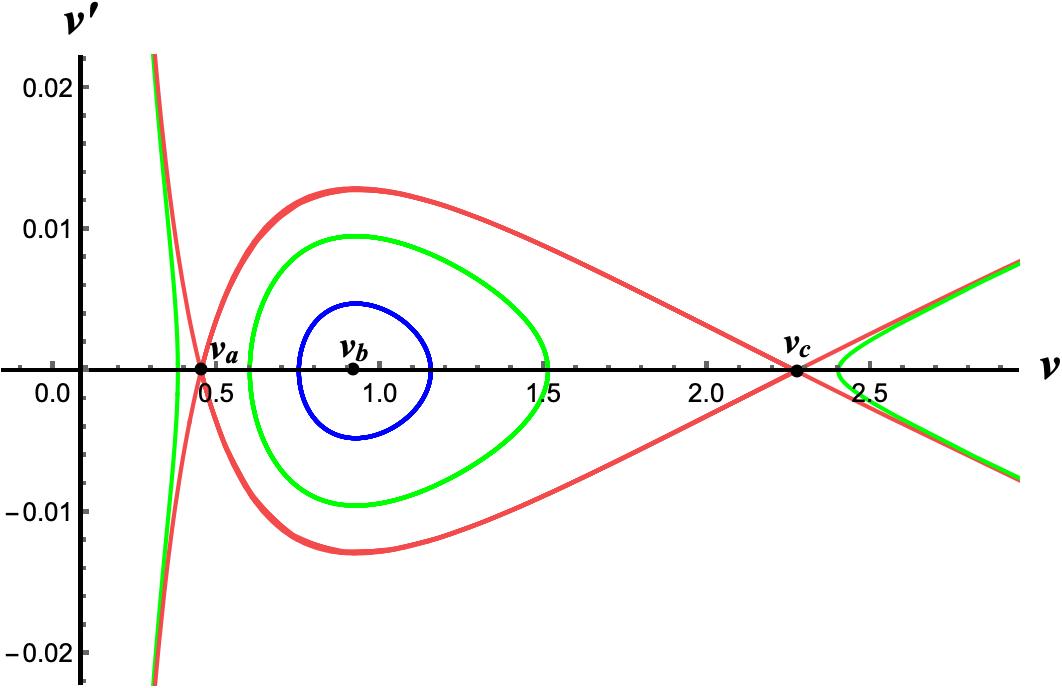}
  \caption{$P-v$ isotherm (the top row) and the corresponding $v'-v$ phase portrait (the bottom row) for three cases. In Case 1 (left), we set $B=0.064$. There is a homoclinic orbit connecting $v_{c}$ to itself (the green line). In Case 2 (middle), we set $B=0.05$. There exists a homoclinic orbit connecting $v_{a}$ to itself (the green line). In Case 3 (right), we set $B=P_0=0.0588359$. There is a heteroclinic orbit connecting $v_{a}$ to $v_{c}$ (the red line). Other parameters are set as $\beta=0.1$, $a=0.1$, $\omega=-2/3$ and $T_0=0.8T_{c}=0.201051$. The dark dashed line is obtained under Maxwell's equal area law.}\label{fig3.1}
  \end{center}
\end{figure*}

There are many kinds of the ambient pressures $B$ we can choose.
Particularly in our research, we only consider the following three cases:
\begin{enumerate}[ {\rm Case} 1 ]
    \item $B$ is set in the range $P(v_2,T_0)<B<P(v_\b,T_0)$. We get a homoclinic orbit connecting saddle point $v_c$ to itself. The portrait of the unperturbed equation in the $v-v'$ phase plane is displayed at the lower-left corner in Fig. \ref{fig3.1}.
    \item $B$ is set in the range $P(v_\a,T_0)<B<P(v_2,T_0)$. We get a homoclinic orbit connecting saddle point $v_a$ to itself. The portrait in the $v-v'$ phase plane is plotted at the lower-middle part in Fig. \ref{fig3.1}.
    \item $B$ is set in the range $B=P(v_2,T_0)$. We get a  heteroclinic orbit connecting $v_a$ to $v_c$. The portrait in the $v-v'$ phase plane is shown at the lower-right corner in Fig. \ref{fig3.1}.
\end{enumerate}

In all the three cases, $v_a$, $v_c$ are saddles, and $v_b$ is a center.
Each of the $v'-v$ panels has a homoclinic or heteroclinic passing through saddle point, which makes it convenient for us to switch on the spatially periodic perturbation and to depict the chaotic behavior in phase space.

Now, we commence to turn on the perturbation to find what will happen on the equilibrium configuration of the quintessential Bardeen-AdS black hole extended phase space.
For simplicity but without loss of generality, we consider the first hydrodynamical mode.
So the dynamical equation becomes
\begin{equation}\label{12}
    A\frac{d^2v}{dx^2}=B-P\left(v,T\right)-\frac{\epsilon\cos\left(sx\right)}{v}\,.
\end{equation}

For a given dynamical system $\dot{y}=f\left(y\right)+\epsilon g\left(y,t\right)\,,\,y\in \mathbb{R}^{2m}$, the Melnikov function can be calculated by \cite{Melnikov1963s}
\begin{equation}
  \begin{split}
      \mathbf{M}\left(t_{\epsilon}\right)=&\int_{-\infty}^{+\infty}\{\mathcal{H}_{0},\mathcal{H}_{1}\}\left(y(t),\dot{y}(t),t+t_{\epsilon}\right)dt\\
      =&\int_{-\infty}^{+\infty}f^{T}\left(y_{0}\left(t-t_{\epsilon}\right)\right)\mathbf{J}_{m=1}g\left(y_{0}\left(t-t_{\epsilon}\right),t\right)dt\,,
  \end{split}
\end{equation}
where the subscript $m$ represents the number of degrees of freedom and $\mathbf{J}_{m=1}$ is a $2\times 2$ matrix given by
\begin{equation}
  \mathbf{J}_{m=1}=
  \begin{pmatrix}
      0 & 1 \\ -1 & 0
  \end{pmatrix}\,.
\end{equation}
$\mathcal{H}_{0}\left(y,\dot{y}\right)$ is the completely integrable part of the total Hamiltonian and $\mathcal{H}_{1}\left(y,\dot{y},t\right)$ denotes the periodically perturbed part.

Given that the unperturbed system $\mathcal{H}_{0}\left(y,\dot{y}\right)$ possesses a hyperbolic fixed point $\left(y_{0},\dot{y}_{0}\right)$ and a homoclinic orbit $\left(y_{0}\left(t\right),\dot{y}_{0}\left(t\right)\right)$ in the phase panel $\left\{y,\dot{y}\right\}$, we can naturally define the stable/unstable manifold: A set of all the points on the homoclinic orbit $\left(y_{0}\left(t\right),\dot{y}_{0}\left(t\right)\right)$ approaches the fixed point $\left(y_{0},\dot{y}_{0}\right)$ in the limit of $t\to+\infty/t\to-\infty$.
Hereafter, they are written as $\ {\rm W}^{s}/{\rm W}^{u}$ respectively.
Similarly, for a heteroclinic orbit $\left(y_{12}\left(t\right),\dot{y}_{12}\left(t\right)\right)$ which is obtained through two hyperbolic fixed points $\left(y_{1},\dot{y}_{1}\right)$ and $\left(y_{2},\dot{y}_{2}\right)$, we have two types of the stable/unstable manifolds $\ {\rm W}^{s}_{1}/{\rm W}^{u}_{1}$ and $\ {\rm W}^{s}_{2}/{\rm W}^{u}_{2}$ respectively.
Under unperturbed circumstances, the stable and unstable manifolds coincide along the homoclinic/heteroclinic orbit.
If we turn on the perturbation, the stable and unstable manifolds will be separated by a distance $d\left(t_{\epsilon}\right)$.
Since the perturbed Hamiltonian is approximately integrable, we can securely calculate $d\left(t_{\epsilon}\right)$ in the perpendicular direction of the original unperturbed homoclinic orbit. Therefore, one can approximately give
\begin{equation}
	d\left(t_{\epsilon}\right)\sim\frac{\epsilon\mathbf{M}\left(t_{\epsilon}\right)}{{\rm D}}+\mathcal{O}\left(\epsilon^{2}\right)\,,
\end{equation}
where ${\rm D}$ denotes some non-zero functions defined in unperturbed system.
It is obvious that once the Melnikov integral $\mathbf{M}\left(t_{\epsilon}\right)$ has a simple zero, the transversal intersection of the stable and unstable manifolds will emerge, i.e. the homoclinic/heteroclinic tangle, which mathematically represents a Smale horseshoe in the phase space, and will exhibit chaotic properties.

Thinking of Eq.(\ref{12}), we calculate the Melnikov function with $x$ taking the part of the integral variable.
So we have
\begin{equation}\label{15}
        \mathbf{M}(x_0)=\int^{+\infty}_{-\infty}-\frac{\dot{v}_{0}\left(x-x_0\right) \cos \left(sx\right)}{v_0\left(x-x_0\right)}{\rm d}x\,,
\end{equation}
where $\left(v_0(x)\,,\dot{v}_0(x)\right)$ is the homoclinic orbit for Case 1 and Case 2, and the heteroclinic orbit for Case 3.
Then we can rewrite Eq.(\ref{15}) as
\begin{equation}
    \mathbf{M}(x_0)=-L\cos\left(sx_0\right)+N\sin\left(sx_0\right)
\end{equation}
with
\begin{equation}
    \begin{aligned}
        L=&\int^{+\infty}_{-\infty}\frac{\dot{v}_0\left(x-x_0\right)\cos\left(s\left(x-x_0\right)\right)}{v_0\left(x-x_0\right)}{\rm d}\left(x-x_0\right)\,,\\
        N=&\int^{+\infty}_{-\infty}\frac{\dot{v}_0\left(x-x_0\right)\sin\left(s\left(x-x_0\right)\right)}{v_0\left(x-x_0\right)}{\rm d}\left(x-x_0\right)\,.
    \end{aligned}
\end{equation}
If $L=0$ and $N=0$, $\mathbf{M}(x_0)$ is equal to zero;
If $L\neq 0$ and $N=0$, $\mathbf{M}(x_0)$ has a simple zero at $sx_0=\left(2k+1\right)\pi/2$ with $m\in \mathbb{Z}$; If $L=0$ and $N\neq 0$, $\mathbf{M}(x_0)$ has a simple zero at $sx_0=k\pi$ with $m\in \mathbb{Z}$;
If $L\neq 0$ and $N\neq 0$, $\mathbf{M}(x_0)$ has a simple zero at $sx_0=\arctan \left(L/N\right)$.
Therefore, the Melnikov function possesses simple zeros in any instances.
So there will always be thermal chaos on the equilibrium configuration of the QB-fluid even under a tiny spatial perturbation as Eq.(\ref{10}).
\begin{figure*}[t]
  \begin{center}
\includegraphics[width=0.26\textwidth,height=0.15\textheight]{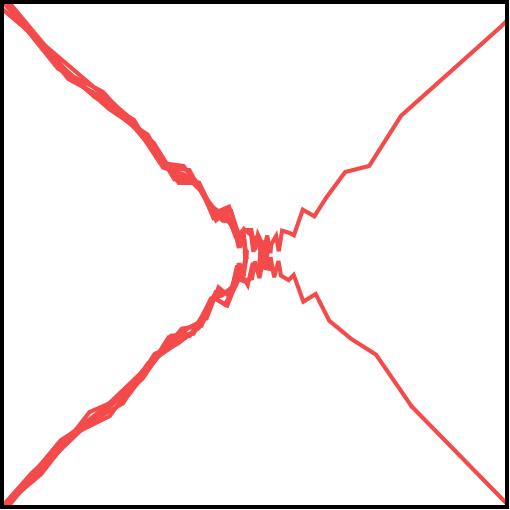}
  \includegraphics[width=0.26\textwidth,height=0.15\textheight]{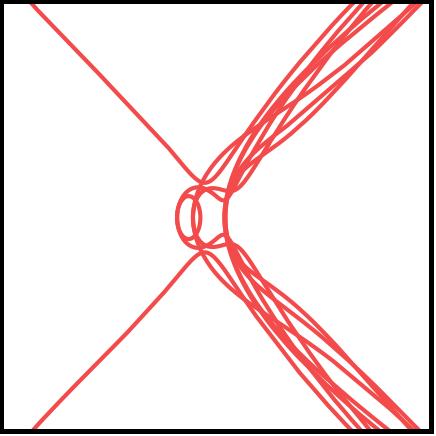}
  \includegraphics[width=0.26\textwidth,height=0.15\textheight]{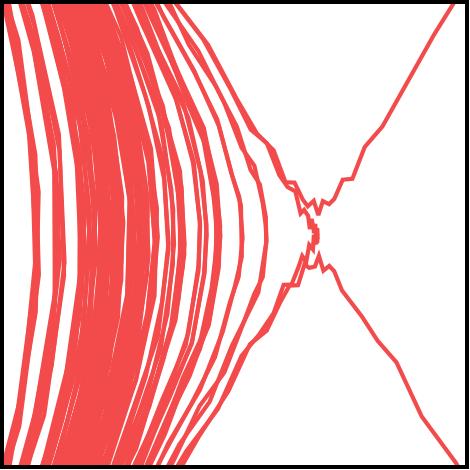}
  \caption{Local chaotic behavior of the homoclinic/heteroclinic orbits connecting saddle points originally shown respectively in the bottom row of Fig.\ref{fig3.1} for 3 cases in proper order. The parameters are setting as $\beta=0.1$, $a=0.1$, $\omega=-2/3$, $\epsilon=0.001$, $s=0.1$ and $\Omega=0.1$.}\label{fig3.2}
  \end{center}
\end{figure*}

We check the local properties around the saddle points in the aforementioned three cases in Fig.\ref{fig3.2}.
The phase planes exhibit complex and strange shapes that are actually the homoclinic/heteroclinic tangles, i.e. the embodiment of chaos.

\section{Temporal Chaos Arising in the Spinodal Region}\label{sec4}
Now we consider how the QB-fluid system evolves in the spinodal region at $T_0<T_c$ under a temporally periodic perturbation
\begin{equation}
  T=T_0+\epsilon\gamma \cos(\Omega t)\cos(M)\,,
\end{equation}
where $\Omega$ is the fluctuation angular frequency and $\gamma$ can be regarded as the relative amplitude of the perturbation associated with a small viscosity of the quintessential Bardeen-AdS black hole in the extended phase space.
The spinodal region denotes the unstable region which is mathematically bounded by the local minimum and local maximum of the isotherm as shown in Fig.\ref{fig2.1} with a pink line.

With Eq.(\ref{9}), regarding $x$ as a function of $M$ and $t$, we define
\begin{equation}
  \begin{aligned}
      \frac{{\rm d}x(M,t)}{{\rm d}M}:=&\rho^{-1}\equiv v\,,\\
      \frac{{\rm d}x(M,t)}{{\rm d}t}:=&u\,,
  \end{aligned}
\end{equation}
where $v$ is just the specific volume mentioned in Section \ref{sec2}, and $u$ denotes the velocity.

Therefore, the balance equations can be expressed in terms of $\left\{u,v\right\}$ as
\begin{equation}\label{21}
    \left\{  
        \begin{aligned}
            \frac{{\rm d}v}{{\rm d}t}= \frac{{\rm d}u}{{\rm d}M}\,,&  \\  
        \frac{{\rm d}u}{{\rm d}t}= \frac{{\rm d}\mathbf{T}}{{\rm d}M}\,.& \\
        \end{aligned}
    \right.
\end{equation}
In the temporal evolving process, the Piola stress tensor $\mathbf{T}$ given in the van der Waals-Korteweg theory of capillarity is
\begin{equation}\label{22}
  \mathbf{T}=-P(v,T)+\mu \frac{{\rm d}u}{{\rm d}M}-A\frac{{\rm d}^2v}{{\rm d}M^2}\,,
\end{equation}
where $\mu\equiv \epsilon\mu_{0}$ is the small viscosity of the QB-fluid.
Substituting Eq.(\ref{22}) into Eq.(\ref{21}) and rewriting the variable with $M\to sM,t\to st,x\to sx,\mu\to \epsilon\mu_{0}$, we have
\begin{equation}\label{23}
  \frac{{\rm d}^2x}{{\rm d}t^2}=-\frac{{\rm d}P}{{\rm d}M}+ \epsilon\mu_{0}s\frac{{\rm d}^3x}{{\rm d}t{\rm d}M^2}-As^2\frac{{\rm d}^4x}{{\rm d}M^4}\,.
\end{equation}

\begin{figure*}[h]
  \begin{center}
\includegraphics[width=0.4\textwidth,height=0.2\textheight]{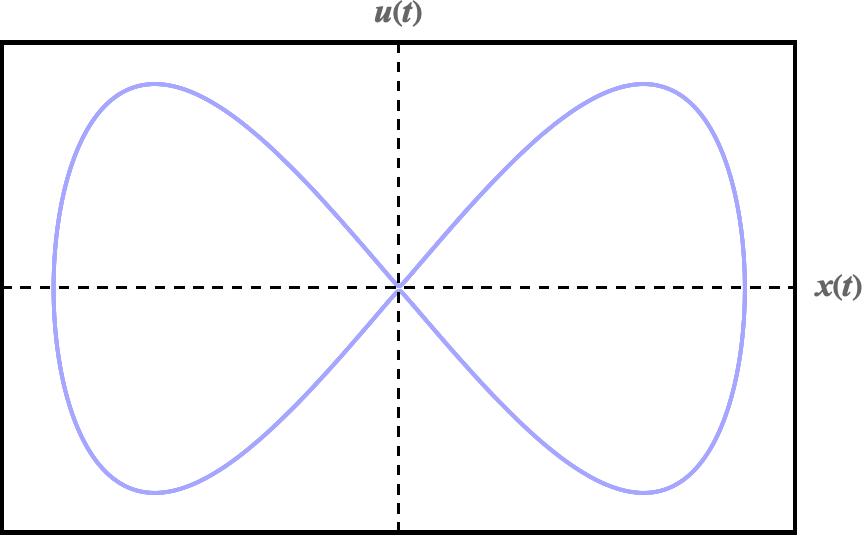}
\includegraphics[width=0.4\textwidth,height=0.2\textheight]{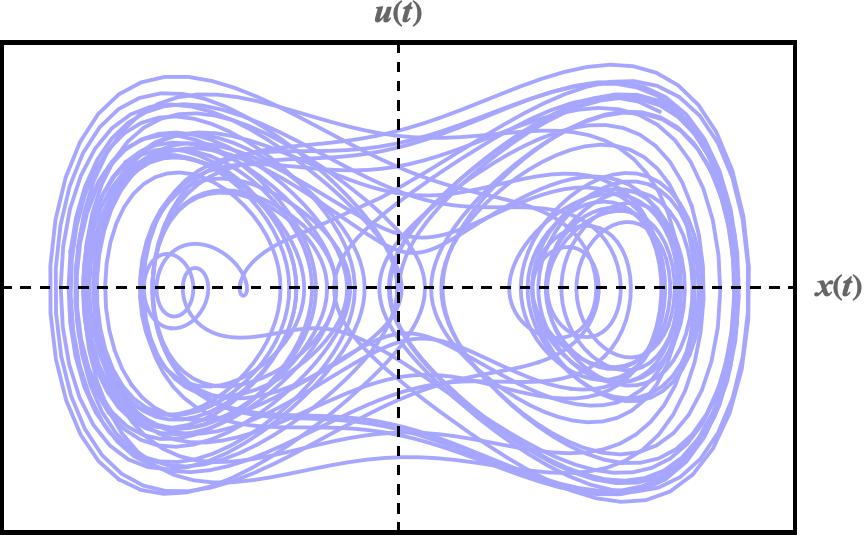}
  \caption{The evolution of the unperturbed equations (left) and perturbed equations for $\gamma>\gamma_{c}=0.000012641$ (right) with fixed $T_0=0.8T_{c}$ and $v_{0}=0.687972$ . We set $\beta=0.1$, $\Omega=0.1$, $\epsilon=0.001$, $A= 0.2$, $s=0.001$, $a=0.1$, $\omega=-2/3$ and $\mu_{0}=0.1$.}\label{fig3.3}
  \end{center}
\end{figure*}

Expanding $P\left(v,T\right)$ about the saddle point $\left(v_0,T_0\right)$ given by ${\rm d}^2P/{\rm d}v^2=0$ in the spinodal region to truncate at cubic terms and taking the first hydrodynamical mode $v\sim v_0+x\left(t\right) \cos M$ and $u\sim u\left(t\right) \sin M$, we solve the dynamical equation (\ref{23}) as
\begin{equation}\label{24}
    \begin{aligned}
        \dot{x}=&u\,,\\
        \dot{u}=&\left(P_{v}\left(v_0,T_0\right)-As^2\right)+\epsilon\gamma \cos\left(\Omega t\right)\left(P_{T}\left(v_0,T_0\right)+\frac{3P_{vvT}\left(v_0,T_0\right)}{8}x^2\right)\\
        &+\frac{P_{vvv}\left(v_0,T_0\right)}{8}x^3-\epsilon\mu_0 s u\,.
    \end{aligned}
\end{equation}
Here, dot denotes derivation with respect to $t$ and subscript represents derivation with respect to $v$ or $T$.
Rewriting Eq.(\ref{24}) as 
$\dot{y}=f\left(y\right)+\epsilon g\left(y,t\right)\,,\,y\in \mathbb{R}^{2m}$ and defining $z=\left(x,u\right)^T$, we have
\begin{equation}
    \begin{aligned}
        f(z)=& 
        \begin{pmatrix}
            u \\  P_{v}\left(v_0,T_0\right)-As^2
        \end{pmatrix}\,,\\
        g(z)=&
        \begin{pmatrix}
            0 \\  \gamma \cos\left(\Omega t\right)\left(P_{T}\left(v_0,T_0\right)+\frac{3P_{vvT}\left(v_0,T_0\right)}{8}x^2\right)-\mu_0 s u
        \end{pmatrix}\,.
    \end{aligned}
\end{equation}
We first turn off the perturbation $\left(\epsilon=0\right)$.
The analytical solution can be obtained
\begin{equation}
    \begin{aligned}
    z_0(t)=&
    \begin{pmatrix}
        x_0(t) \\  u_0(t)
    \end{pmatrix}\\
    =&
    \begin{pmatrix}
        \pm\frac{4P_{v}\left(v_0,T_0\right)-As^2}{\left(-P_{vvv}\left(v_0,T_0\right)\right)^{1/2}}{\rm sech} \left(\left(P_{v}\left(v_0,T_0\right)-As^2\right)t\right)  \\ 
        \mp \frac{\left(2P_{v}\left(v_0,T_0\right)-As^2\right)^2}{\left(-P_{vvv}\left(v_0,T_0\right)\right)^{1/2}}{\rm sech} \left(\left(P_{v}\left(v_0,T_0\right)-As^2\right)t\right){\rm tanh} \left(\left(P_{v}\left(v_0,T_0\right)-As^2\right)t\right)
    \end{pmatrix}\,.
    \end{aligned}
\end{equation}
It is obvious that there is a homoclinic orbit connecting to the saddle point $(0,0)$ which is shown on the left phase panel of Fig.\ref{fig3.3}.

Now we switch on the temporal perturbation $\left(\epsilon>0\right)$ and detect the existence of chaos utilizing the Melnikov Method.
Substituting Eq. (\ref{22}) into Eq. (\ref{3}), we obtain
\begin{equation}
    \begin{aligned}
    \mathbf{M}(t_\epsilon)
    =&\int^{+\infty}_{-\infty}{\rm d}t \frac{\left(2P_{v}\left(v_0,T_0\right)-As^2\right)^2}{\left(-P_{vvv}\left(v_0,T_0\right)\right)^{1/2}}\gamma\cos\left(\Omega t\right)\\
    &\times \left[\frac{6\left(P_{v}\left(v_0,T_0\right)-As^2\right)^2P_{vvT}\left(v_0,T_0\right)}{P_{vvv}\left(v_0,T_0\right)}{\rm sech}^2 \left(\left(P_{v}\left(v_0,T_0\right)-As^2\right)(t-t_\epsilon)\right)-P_T\left(v_0,T_0\right)\right]\\
    &\times {\rm sech} \left(\left(P_{v}\left(v_0,T_0\right)-As^2\right)(t-t_\epsilon)\right){\rm tanh}\left(\left(P_{v}\left(v_0,T_0\right)-As^2\right)(t-t_\epsilon)\right)
       \\ 
        &+ 
        \int^{+\infty}_{-\infty}dt \frac{16\mu_0 s \left(2P_{v}\left(v_0,T_0\right)-As^2\right)^4}{P_{vvv}\left(v_0,T_0\right)}{\rm sech}^2 \left(\left(P_{v}\left(v_0,T_0\right)-As^2\right)(t-t_\epsilon)\right)\\
        &\times {\rm tanh}^{2}\left(\left(P_{v}\left(v_0,T_0\right)-As^2\right)(t-t_\epsilon)\right)\\
        \equiv&
        \gamma\Omega \mathbf{R}_1 \sin\left(\Omega t_\epsilon\right)+\mu_0 s \mathbf{R}_2
    \end{aligned}
\end{equation}
with 
\begin{equation}
    \begin{aligned}
        \mathbf{R}_1\equiv&\frac{8\pi}{\left(P_{vvv}\left(v_0,T_0\right)\right)^{1/2}}\left[P_{vT}\left(v_0,T_0\right)-\frac{P_{vvT}\left(v_0,T_0\right)}{P_{vvv}\left(v_0,T_0\right)}\left(\Omega^2+\left(P_{v}\left(v_0,T_0\right)-As^2)\right)^{2}\right)\right]\\
        &\times \frac{{\rm exp}\left[\frac{\pi \Omega}{2\left(P_{v}\left(v_0,T_0\right)-As^2)\right)}\right]}{1+{\rm exp}\left[\frac{\pi \Omega}{2\left(P_{v}\left(v_0,T_0\right)-As^2)\right)}\right]}\,,\\
        \mathbf{R}_2\equiv&\frac{32\left(P_{v}\left(v_0,T_0\right)-As^2)\right)^3}{3P_{vvv}\left(v_0,T_0\right)}\,.
	\end{aligned}
\end{equation}
It is not difficult to check that $\mathbf{M(t_\epsilon)}$ will possess a simple zero only if the assumed perturbed amplitude $\gamma$ is larger than a critical value $\gamma_c$, i.e.
\begin{equation}
    \gamma \ge \gamma_c \equiv \lvert \frac{s \mu_0 \mathbf{R}_2}{\Omega \mathbf{R}_1} \rvert\,.
\end{equation}
The perturbed dynamic system is solved numerically on the right panel of Fig.\ref{fig3.3}.
If we set $\gamma<\gamma_{c}$, one can easily see the original homoclinic orbit staying still on the phase panel as before.
But if the value of $\gamma$ becomes larger than $\gamma_{c}$, the evolutionary process will be extremely complex and unpredictable, which means that the chaotic phenomenon appears in the dynamical system.

\begin{figure*}
  \begin{center}
  \includegraphics[width=0.4\textwidth,height=0.2\textheight]{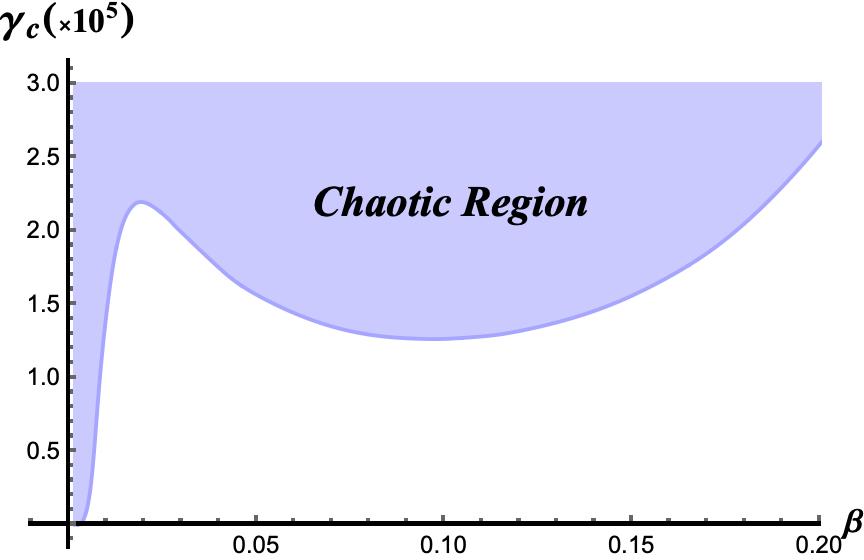}
  \includegraphics[width=0.44\textwidth,height=0.2\textheight]{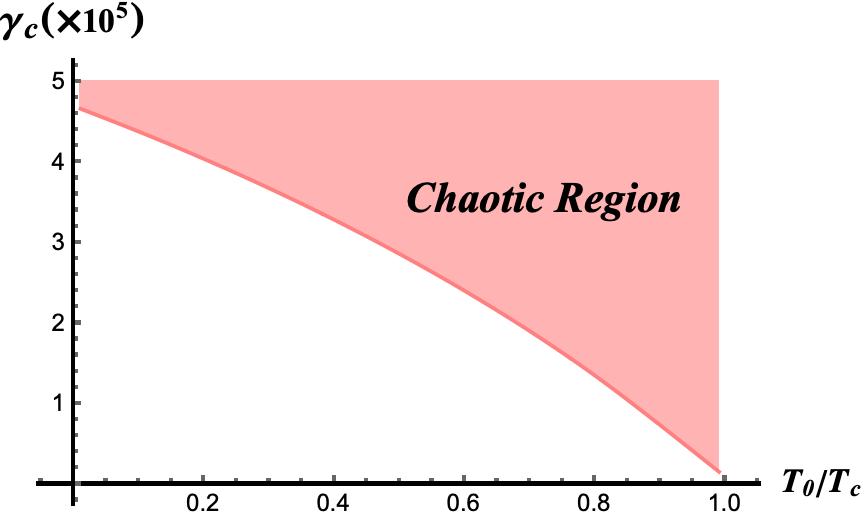}
  \caption{Dependence of the critical amplitude $\gamma_{c}$ on magnetic monopole charge $\beta$ under the condition of $T_0=0.8T_{c}$ (left).
  The shaded region denotes the existence of chaos.
  Dependence of the critical amplitude $\gamma_{c}$ on the initial temperature $T_0$ (right).
  Other parameters are set as $\Omega=0.1$, $\epsilon=0.001$, $A= 0.2$, $s=0.001$, $a=0.1$, $\omega=-2/3$, $\mu_{0}=0.1$.}\label{fig3.4}
  \end{center}
\end{figure*}

The dependences of $\gamma_c$ on the non-linear electrodynamics (NLED), the initial conditions and the quintessence dark energy are worth investigating in detail.

Firstly, as is shown in Fig.\ref{fig3.4}, with the increase of magnetic monopole charge $\beta$, the critical amplitude $\gamma_c$ of temporally periodic perturbation first increases to a local maximum, then decreases to a minimum, and finally increases again.
The upward slope of the $\gamma_c-\beta$ curve (critical curve) is very large for $\beta\in\left[0.20,\infty\right)$, so it is hard to exhibit the chaotic behavior for the perturbed B-flow with a large $\beta$.
In the fuzzy region of $\beta\in\left[0.01,0.20\right]$, there is a maximum value at $\beta\approx 0.0226$ and a minimum value at $\beta\approx 0.094$.
We find that in a very small $\beta$ region, chaotic phenomenon appears more easily.
It is corresponding to the fact that the QB-fluid with a small $\beta$ is too weak to bear a strong enough fluctuation.
We believe that it is the cause of two bends of the critical curve.

Secondly, we investigate the dependence of the critical amplitude $\gamma_c$ on the initial temperature $T_0$ on the right panel of Fig.\ref{fig3.4}.
It is shown that the critical amplitude $\gamma_c$ is in negative correlation to the temperature ratio $T_0/T_c$.
It is probably because once we set the initial temperature $T_0$ closer to the critical temperature $T_c$, the $P-v$ oscillation will get weaker and hence the unstable spinodal region becomes narrower.
As a consequence, the QB-fluid can not remain stable even after we impose a tiny perturbation.

\begin{figure*}
  \begin{center}
      \includegraphics[width=0.4\textwidth,height=0.2\textheight]{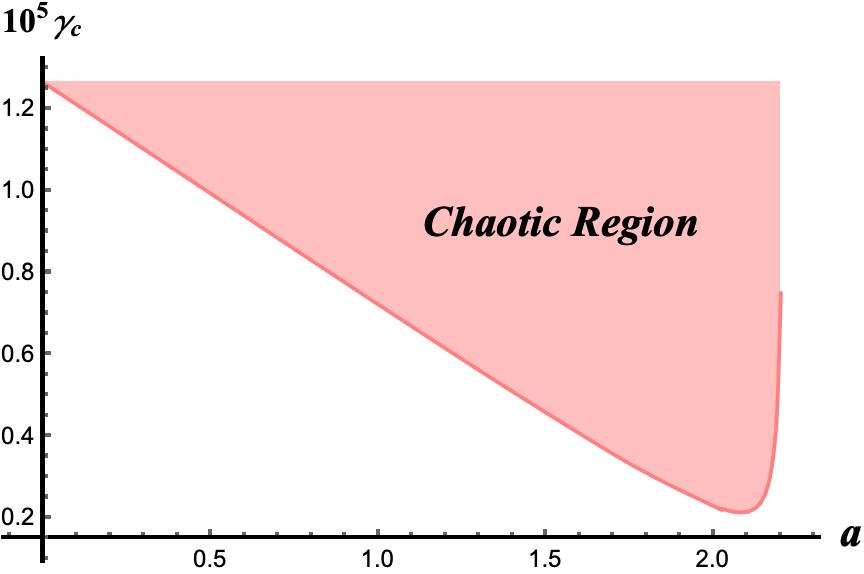}
      \includegraphics[width=0.4\textwidth,height=0.2\textheight]{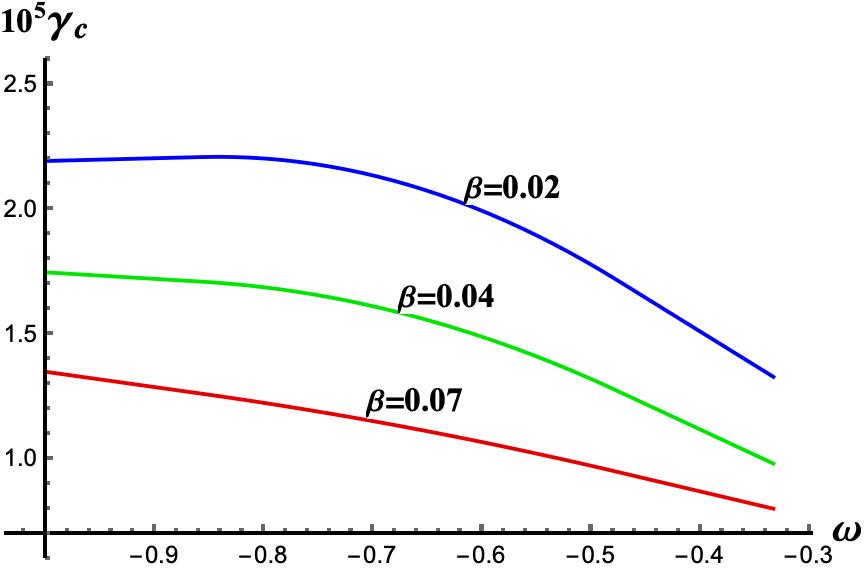}
  \caption{Dependence of the critical amplitude $\gamma_{c}$ on quintessence normalization parameter $a$ (left) with $T_0=0.8T_{c}$, $\omega=-2/3$, $\epsilon=0.001$, $A= 0.2$, $s=0.001$ and $\mu_{0}=0.1$.
  Dependence of the critical amplitude $\gamma_{c}$ on quintessence state parameter $\omega$ (right) for the quintessential Bardeen-AdS black hole at different $\beta$ with $T_0=0.8T_{c}$, $a=0.5$, $\epsilon=0.001$, $A= 0.2$, $s=0.001$ and $\mu_{0}=0.1$.}\label{fig3.5}
  \end{center}
\end{figure*}

Thirdly, we want to find how the quintessence normalization parameter $a$ and the quintessence state parameter $\omega$ affect the critical amplitude $\gamma_c$.
It is shown in Fig.\ref{fig3.5} that the quintessence normalization parameter $a$ has a non-trivial impact on the existing condition of temporal chaos.
The value of $\gamma_c$ shows an approximately linear decline before reaching the global minimum at $a_{c}\approx 2.11$, but will tend to positive infinity rapidly until the quintessence normalization parameter $a$ meets an upper bound $a_{{\rm max}}$ which is restricted by the non-extreme condition of quintessential Bardeen-AdS black holes.
Besides, we find that on the right panel, with $\omega$ varying from $-1$ to $-1/3$, the value of $\gamma_c$ decreases more and more rapidly.
In effect, accroding to the relation $\rho_{q}=-3a\omega/2r^{3\left(\omega+1\right)}$, the raising values of $a$ and $\omega$ directly lead to the increase in the quintessence dark energy density.
Therefore, combining the above two factors, we find that there exists a critical value $\rho_c$ of the quintessence dark energy density $\rho$, below which the QB-fluid is less bearable to the temporally periodic fluctuation and thus easier to exhibit chaotic behavior in the spinodal region.
But the reverse situation applies when the quintessence acquires an energy level that is larger than $\rho_c$.
Because the response to thermal fluctuation is intrinsically linked with the small viscosity $\mu$ of the QB-fluid, the critical amplitude $\gamma_c$ should be a positive function of $\mu$.
The change of $\gamma_c$ in fact drops a hint about the variation of $\mu$.
Therefore, we can audaciously make a conjecture that the effect of quintessence dark energy appears quite similar to an enhancing/damped mechanism.
In the region of the energy density $\rho<\rho_c$, the existence of quintessence leads to a reduction in the viscosity of the QB-fluid, to wit an enhancement in conversing the total energy during the process of passive flow caused by the temporally periodic perturbation.
Thereby, in this case, the QB-fluid behaves more like a perfect fluid and thus is more possible to exhibit chaotic behavior.
Conversely, given the energy density $\rho>\rho_c$, the system acquires a higher viscosity so that it is endowed with the ability of enduring a larger thermal fluctuation, which is analogous to a damped mechanism.

\section{Conclusions}\label{sec5}

We have investigated the existing conditions and properties of thermal chaos under
temporally/ spatially periodic perturbations in the extended phase space of quintessential Bardeen-AdS black holes, namely the QB-fluid.

Imposing a spatially periodic perturbation on the equilibrium configuration, we find that the spatial chaos appears even under a tiny perturbed amplitude.
The similar result has also been found in the van der Waals fluid \cite{Polcar:1985} and other black hole solutions in Refs. \cite{Chabab:2018lzf,Mahish:2019tgv,Chen:2019bwt,Dai:2020wny}.
But to be specific, the chaotic features of the $v'-v$ phase space that is plotted in Fig.\ref{fig3.2} looks distinct from other systems.
There exist unusual shapes of the homoclinic/heteroclinic tangles around saddle points.

While considering a temporally periodic perturbation imposed in the unstable spinodal region of the QB-fluid, it is found that the phase space exhibits the chaotic behavior only if the perturbed amplitude $\gamma$ is larger than a critical value $\gamma_c$.
The value of $\gamma_c$ mainly depends on the NLED (specificly the magnetic monopole charge $\beta$), the initial temperature $T_0$ and the surrounding quintessence dark energy.

Firstly, a quite large $\beta$ will make it difficult for chaos in the $v'-v$ phase panel.
But in the region of smaller $\beta$s, $\g_c$ first increases to a local maximum, then decreases to a minimum, and finally increases again rapidly.
In the smaller $\beta$ region, the thermal phase transition is too weak to bear a relatively strong fluctuation, and the metric function itself depends on $\beta$ sensitively and complicatedly.
By contrast, in the investigation of other AdS black hole extended phase spaces, the critical amplitude $\g_c$ is a monotonous decrease function of the charge $Q$ \cite{Chabab:2018lzf,Mahish:2019tgv,Chen:2019bwt,Dai:2020wny}.
It may drop a hint that the particularity of NLED leads to the two extreme points of the $\g_c-\beta$ curve in Fig.\ref{fig3.3}.
Hence, it is natural to infer that the $\g_c-\beta$ curve might also show the same trend in consideration of other regular black hole solutions of Einstein's field equation coupling with a non-linear electromagnetic field.
As it turns out, through our simple calculation, this conjecture is valid for a Hayward black hole and other new regular black hole classes.

Secondly, it is found that $\g_c$ is in negative correlation to the ratio $T_0/T_c$, which means that if we originally immerse the QB-fluid tube into a "heat bath" with a lower initial temperature $T_0$, we will find it more difficult for chaos.

Thirdly, considering the quintessence dark energy, we conclude that its effect appears quite similar to an enhancing/damped mechanism.
There exists a critical value $\rho_c$ of the quintessence dark energy density $\rho$.
In the region of the energy density $\rho<\rho_c$, the existence of quintessence leads to a reduction in the viscosity of the QB-fluid, to wit an enhancement in conversing the total energy during the process of passive flow caused by the temporally periodic perturbation.
Thereby, the QB-fluid behaves more like a perfect fluid and thus is more possible to exhibit chaotic behavior.
Conversely, given the energy density $\rho>\rho_c$, the system acquires a higher viscosity so that it is endowed with the ability of enduring a larger thermal fluctuation, which is analogous to a damped mechanism.

The results we have found do not only reflect the non-trivial influence that NLED and the quintessence dark energy have upon the black hole both thermodynamically and hydrodynamically, but also in all likelihood, provide us with enlightenment to draw a more reasonable picture about the evolution of black holes due to a thermal fluctuation.
It is hopeful that we can apply our results to illustrate what the occurrence of chaos in the extended phase space exactly means and to describe how black holes evolves in a quintessential environment in further studies.
A terrific idea is to combine our study with the thermodynamic geometry \cite{Ruppeiner:2018pgn} and the stochastic kinetic approach \cite{Li:2020khm}.
Both of them are highly advantageous to investigate black holes under thermal fluctuations.
The former lays particular emphasis on the microscopic interaction while the latter tends to discuss the non-equilibrium evolutionary process.

\section*{Acknowledgement}
This work is supported by the National Natural Science Foundation of China (Grant No. 11235003).

\end{document}